\documentclass[aps, prb, twocolumn, superscriptaddress, preprintnumbers, amsmath, epsFig, floatfix,normalem]{revtex4}
\usepackage{graphicx}
\usepackage{dcolumn}
\usepackage{bm}
\usepackage{ulem}
\usepackage{amsmath}
\usepackage{textcomp}
\usepackage{amssymb}

\usepackage[utf8]{inputenc}
\usepackage[english]{babel}
\usepackage{keyval}
\graphicspath{{Figs/}{Figs:}{\Figs}}
\setkeys{Gin}{width=0.9\columnwidth}

\usepackage[usenames,dvipsnames]{color}
\usepackage[]{ulem}

\newcommand{\comment}[1]{\textcolor{red}{#1}}
\renewcommand{\comment}[1]{\relax}

\newcommand{\todelete}[1]{\textcolor{green}{\sout{#1}}}
\renewcommand{\todelete}[1]{\relax}

\begin{document}

\title{Light-enhanced gating effect at the interface of oxide heterostructure}
\date{\today}
\author{ Neha Wadehra}
\affiliation{Nanoscale Physics and Device Laboratory, Quantum Materials and Devices Unit, Institute of Nano Science and Technology, Knowledge City, Sector-81, S.A.S. Nagar, Mohali, Punjab, 140306, India.}

\author{Ruchi Tomar}
\affiliation{Nanoscale Physics and Device Laboratory, Quantum Materials and Devices Unit, Institute of Nano Science and Technology, Knowledge City, Sector-81, S.A.S. Nagar, Mohali, Punjab, 140306, India.}

\author{Yuichi Yokoyama}
\affiliation{Institute for Solid State Physics, University of Tokyo, Kashiwanoha, Chiba, 277-8581, Japan.}

\author{Akira Yasui}
\affiliation{Japan Synchrotron Radiation Research Institute, Hyogo 679-5198, Japan.}

\author{E. Ikenaga}
\affiliation{Japan Synchrotron Radiation Research Institute, Hyogo 679-5198, Japan.}

\author{H. Wadati}
\affiliation{Institute for Solid State Physics, University of Tokyo, Kashiwanoha, Chiba, 277-8581, Japan.}

\author{Denis Maryenko}
\affiliation{Center for Emergent Matter Science, RIKEN, 2-1 Hirosawa, Wako, Saitama, 351-0198, Japan.}

\author{S. Chakraverty}
\email{suvankar.chakraverty@inst.ac.in}
\affiliation{Nanoscale Physics and Device Laboratory, Quantum Materials and Devices Unit, Institute of Nano Science and Technology, Knowledge City, Sector-81, S.A.S. Nagar, Mohali, Punjab, 140306, India.}

\begin{abstract}
\noindent
In semiconducting materials, electrostatic gating and light illumination are widely used stimuli to tune the electronic properties of the system. Here, we show a significant enhancement of photoresponse at the conducting interface of LaVO$_3$-SrTiO$_3$ under the simultaneous application of light and \textit{negative} gate bias voltage, in comparison to their individual application. On the other hand, the LaVO$_3$-SrTiO$_3$ interface remains largely insensitive to light illumination, when a \textit{positive} gate bias voltage is applied. Our X-ray diffractometer, Raman spectroscopy and photoemission measurements show that unlike the LaAlO$_3$-SrTiO$_3$ interface, migration of oxygen vacancies is not the prime mechanism for the enhanced photoresponse. Rather, we suggest that the photoresponse of our system is intrinsic and this intrinsic mechanism is a complex interplay between band filling, electric field at the interface, strong electron interaction due to mottness of LaVO$_3$ and modification of conducting channel width. 
\end{abstract}
\maketitle

\section{\label{sec:intro}INTRODUCTION}

Over the past few decades, oxides have been playing an important role in the discoveries of emergent phenomena like high temperature superconductivity, multiferroics, colossal magnetoresistance and so on.\cite{Bednorz1986,Cava1988, Reyren2007, Catalan2009, Brinkman2007, Hwang2012,Hasegawa2003,Balal2017,Matsubara2016} The field of oxide based electronic devices received a special attention after the discovery of two dimensional electron gas (2DEG) at the interface of two band insulating oxides namely LaAlO$_3$ (LAO) and SrTiO$_3$ (STO).\cite{Ohtomo2004} This simple interface exhibits wealth of phenomenona including quantum oscillations in conductivity, superconductivity, magnetism, large Rashba spin-orbit coupling, electric field effect, persistance photocurrent etc.\cite{Reyren2007, Caviglia2010, Li2011, Tebano2012,Caviglia2010a}
Many of these properties emerge by controlling the charge carrier density of the system. Among various methods, charge carrier density of this system can be tuned by light illumintaion and electrostatic gating.\cite{Caviglia2008,Tarun2013,Gennaro2015} But unlike semiconductors, such processes in oxides are by far more complicated. Recently, Lei et. al. reported that simultaneous application of light illumination and electrostatic gating affect dramatically the conductivity of the 2DEG hosted at LAO-STO interface. The modulation of interface conductance was explained in terms of "illumination-accelerated interface polarization"  generated from the oxygen deficiency, an extrinsic effect. \cite{Lei2014}

Here, we present experimental evidences suggesting that the photoresponse of 2DEG located at the interface between Mott-insulator LaVO$_3$ (LVO) and band insulator SrTiO$_3$ has intrinsic origin to simultaneous stimulation by light illumination and electrostatic gating. \cite{Hotta2007} As we show later, in contrast to the reported LAO-STO interface, LVO-STO interface exhibits no (or below detectable limit) oxygen deficiency. The dramatic change of interface conductance to both stimuli is a complex process involving band filling, gate induced interface electric field, conducting channel width modification, strong electron-electron interactions and alteration of polar catastrophe condition.

\begin{figure}[t!] \scalebox{1}{\includegraphics{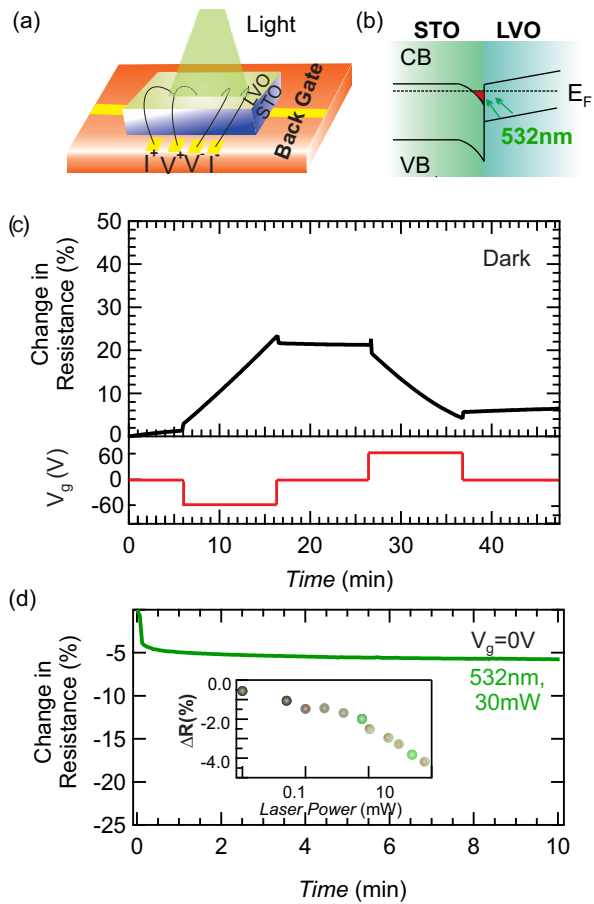}}
\caption{(Color online) (a) Schematic of the measurement setup with back gate applied at the bottom and laser light illuminated from the top of LVO-STO. (b)  Band diagram of LVO-STO depicting excitation of carriers from LVO valence band to quantum well formed on light illumination. (c) Percentage change in resistance of the crystalline LVO-STO kept in dark on application of -60V, 0V, +60V.(d) Percentage change in resistance of the crystalline LVO-STO on illumination with laser light of 532nm wavelength with no bias applied to the back gate.(inset) Percentage change in resistance as a function of laser power after two minutes. All the measurements are performed at room temperature. }
\end{figure}

\section{\label{sec:intro}Results}

Crystalline and amorphous LVO films (hereafter, will be reffered as crystalline and amorphous samples) were grown on TiO$_2$-terminated STO (001) single crystals using pulsed laser depostion (PLD) (see details in Methods). \cite{Hotta2007, Prakash2015} The film thickness was monitored using the reflection high-energy electron diffraction (RHEED) technique and was kept to be 10 monolayer (ML) corresponding to about 4nm. The crystalline LVO-STO sample was conducting down to 2K and showed metallic behavior, whereas the amorphous LVO-STO was insulating even at room temperature suggesting the absence of oxygen vacancies in the sample in contrast to previously reported amorphous LaAlO$_3$-SrTiO$_3$ (LAO-STO) interface. \cite{Lei2014} The temperature dependent resistance of the crystalline sample is shown in supplementary information figure S1. The charge carrier density of the 10ML crystalline LVO-STO evaluated from the Hall measurements done at 300K was 6.6x10$^{13}$cm$^{-2}$ and the corresponding mobility was 1.8cm$^2$V$^{-1}$ s$^{-1}$. The back side of STO substrate was metalized and served as a back gate electrode.    

Firstly, both the amorphous and crystalline samples were back gated in dark and the change in resistance was measured using four probe geometry as shown in Fig. 1(a). The amorphous sample did not show any response to the electrostatic gating. Figure 1(c) shows the percentage change in sheet resistance of the crystalline sample on subsequent application of -60V, 0V and +60V. The percentage in resistance is defined as ((R$_t$-R$_o$)/R$_o$)*100, where R$_o$ is the saturated resistance value of the sample before application of any stimuli and R$_t$ is the resistance value at time t when different stimuli are applied on the samples. The sheet resistance of the sample increased for negative bias and decreased for positive bias. The response of the sample to gating effect was linear in time and the change in resistance on application of both -60V and +60V was about 20\% after 10 minutes.

Next, the photoresponse of samples was checked by illuminating them with laser light at wavelength 532nm and laser power 30mW. No gate voltage was applied to the sample in this experiment. The sheet resistance of the sample upon illumination with laser light was measured as a function of time as shown in Fig. 1(d). While the resistance of crystalline sample changed by 5\%, the amorphous sample did not show any response. Figure 1(d) inset summarizes the laser power dependence of the photoresponse for crystalline sample. The figure shows the value of the percentage change in resistance as a function of laser power after two minutes.

Remarkable results were obtained when the electrostatic gating and light illumination were simultaneously applied. Simultaneous application of negative gate bias and light increased the resistance significantly (Fig. 2(a)). The increase in resistance upon negative gate voltage under light illumination was around 3 times in comparison to that without light. At the same time, the application of  positive back gate voltage had almost no effect on ilumination with light, the resistance decreases by similar amount as it was under light illumination only (Fig. 2(b)). Figure 2(a) shows the change in resistance of the sample on simultaneous application light and negative gate bias upto 10 minutes and then the change in resistance of the sample after turning off the gate bias. Figure 2(b) presents the change in resistance of the sample under simultaneous application light and positive gate bias.

\begin{figure}[t!] \scalebox{1}{\includegraphics{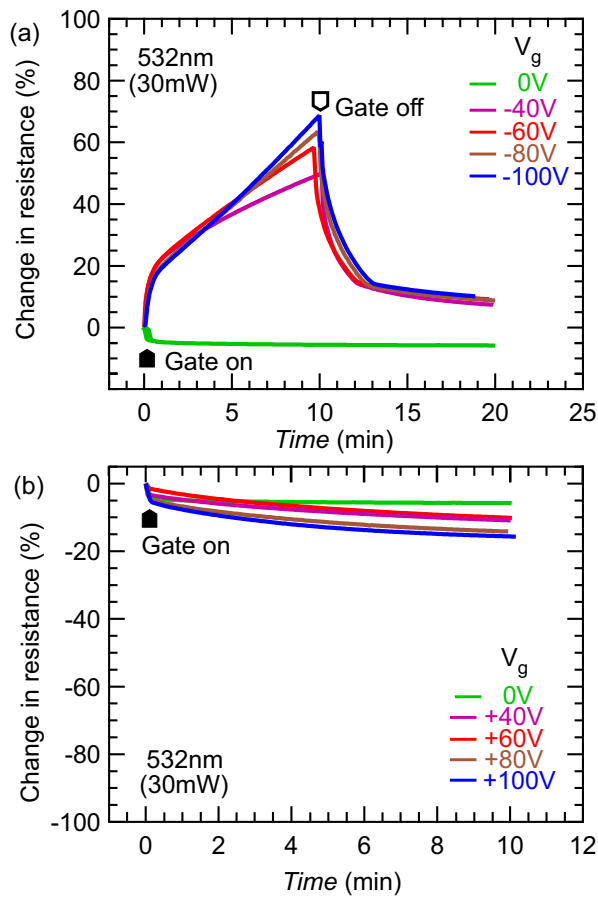}}
\caption{(Color online)  (a) and (b) Room temperature percentage change in resistance of the sample with simultaneous application of 532nm laser light and negative and positive gate bias respectively.}
\end{figure}

\begin{figure*}[t!] \scalebox{2}{\includegraphics{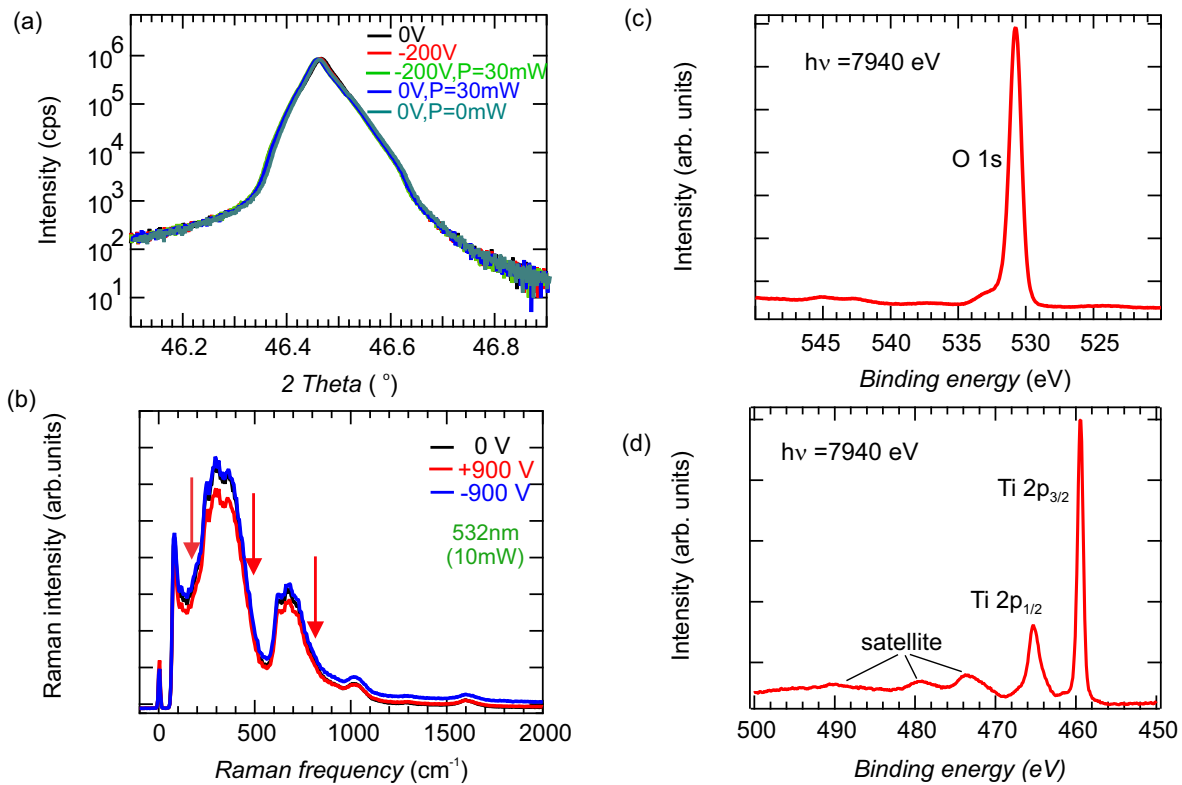}}
\caption{(Color online) (a) XRD data of 002 peak of STO under several scans showing no additional peak with or without the two stimuli applied. All the XRD traces coincide. (b) Raman spectra of crystalline LVO-STO on application of \textpm  900V. Arrows show the position of the additional peaks obtained in ref.[18] on simultaneous application of negative gate and laser light. (c) O 1s core-level HX photoemission spectrum of the thin film. (d) HX photoemission spectrum showing Ti 2p peaks. Peaks at 459eV and 473eV correspond to Ti 2p$_{3/2}$ main and satellite peaks. Peaks at 465eV and 478eV correspond to Ti 2p$_{1/2}$ main and satellite peaks.}  
\end{figure*}

\begin{figure*}[t!] \scalebox{2}{\includegraphics{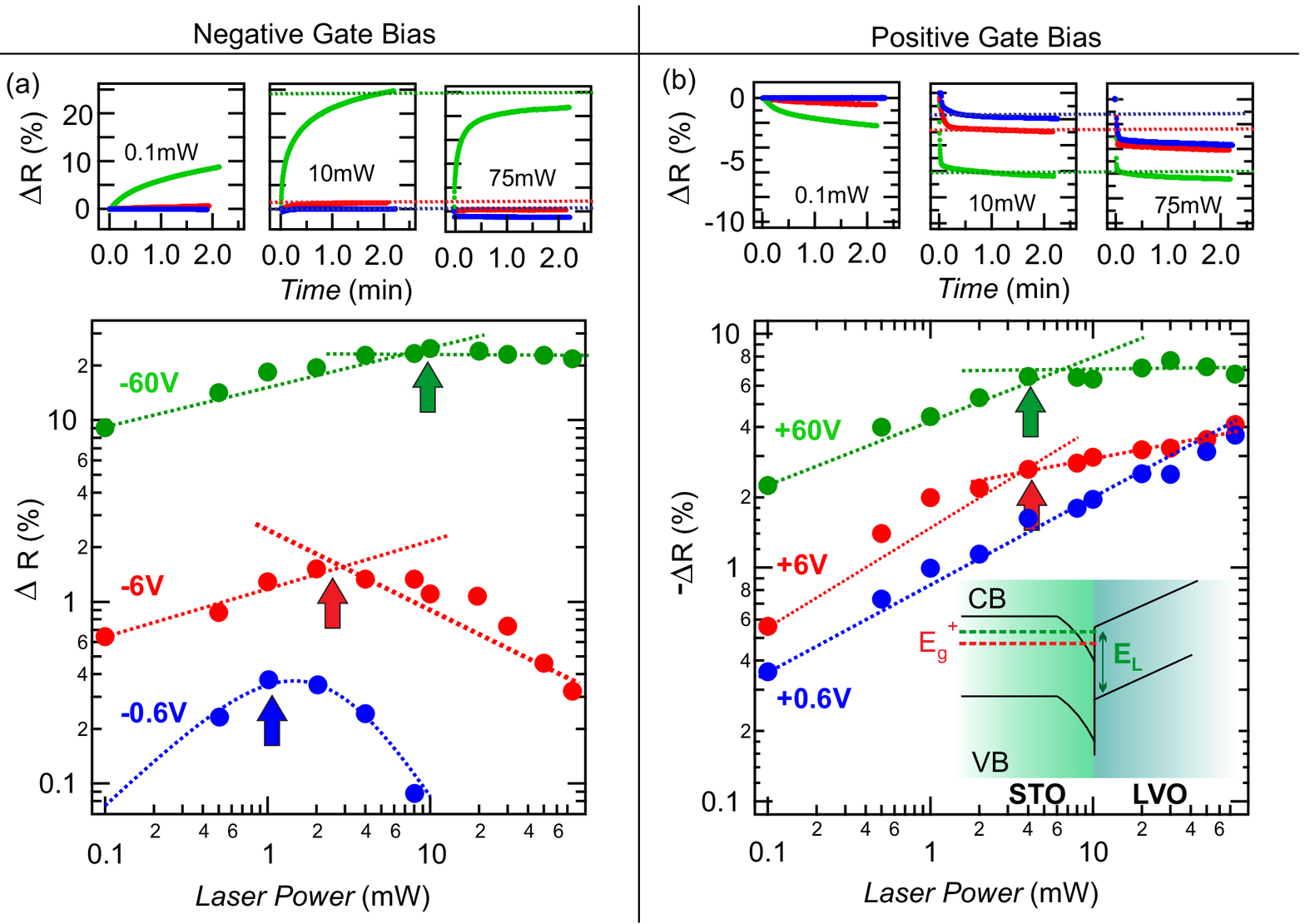}}
\caption{(Color online)(a) and (b) Mode change in resistance of samples on application of negative and positive gate bias along with laser light of varying intensities. The blue markers show change in resistance for \textpm0.6V, red markers show the change in resistance for \textpm6V and the green markers show the resistance change for \textpm60V. Top panel shows the time evolution of the resistance change for 0.1mW, 10mW, 75mW laser intensities in both the cases. (b) (inset) Band diagram of LVO-STO for positive gate bias applied.}
\end{figure*}

Similar phenomenon had been observed in oxygen vacant conducting LAO-STO interface and its origin was ascribed to the gating induced and illumination enhanced lattice polarization. \cite{Lei2014} A mechanism was put forward that the oxygen vacancies created near the interface during the growth of the sample starts migrating away from the interface slowly under the electric field thus turning the interface less conducting. This slow migration is accelerated upon illumination with laser light resulting in an enhanced gating effect. X-ray diffraction and Raman spectroscopy under the application of both stimuli supported the suggested mechanism. A significant change in X-ray pattern of LAO-STO was observed indicating lattice polarization of oxygen vacant STO.

To check for this mechanism in our LVO-STO interface, we performed detailed XRD study using Bruker thin film X-ray diffractometer Discover-8. No significant change/shift in the 002 peak of STO was observed under individually either light or gate or under the application of both simultaneously (Fig. 3(a)). This contrasts to the X-ray studies on LAO-STO and suggests no change in the crystal structure of our sample, indicating no migration of oxygen vacancies. The backgating of LAO-STO and LVO-STO systems cannot deplete the charge carriers at the interface, which could potentially diminish the effect of oxygen migration. In fact, an application of -200V gate voltage changes the charge carrier density by less than 1\% as we show in our supplementary information note 1. Therefore, the electric field is applied across the interface.

To further look into the possible lattice distortion and, hence, the change in the bond length between atoms, under simultaneous application of electrostatic gate and light, Raman spectroscopic studies were done using WITec Confocal Raman Spectrometer. The sample was back gated and the spectra were obtained using 532nm laser light. Figure 3(b) shows the Raman spectra of the crystalline LVO-STO sample with and without gate voltage being applied. One expects the appearance of additional peaks at higher frequencies if the bond length to oxygen changes.\cite{Migoni1976} Such an effect is absent even at \textpm 900V in our sample. This strengthens our conclusion that no lattice distortion is caused by oxygen deficiency migration upon application of gate voltage and illumination with light. The arrows in Fig. 3(b) mark the position of the additional peaks obtained in Raman spectra of the previously reported LAO-STO [18].

X-ray photoemission spectroscopic measurements were performed to detect the presence of oxygen vacancies in our samples. Hard X-ray (HX) photoemission measurements were carried out on the samples at BL-47XU of SPring-8. The HX photoemission spectra were recorded using a Scienta R-4000 electron energy analyzer with a total energy resolution of 300meV at the photon energy of 7.94keV. Before performing these measurements, the sample was kept in dark for sufficiently long time to assure the absence of any effect of applied  stimuli from previous experiments. The position of E$_F$ was determined by measuring the spectra of gold which was in electrical contact with the sample.\cite{Wadati2008} All the spectra were measured at room temperature.  Figure 3(c) shows the O 1s core-level photoemission spectrum of the thin film. The O 1s spectrum consists of a single peak. For oxygen vacancies to be present in the system, there should be additional peaks in the O 1s spectrum at few eV higher binding energy value. In our spectrum, no additional peak or shoulder developing on the higher-binding-energy side of O 1s peak was observed confirming no surface contamination or oxygen vacancy signal in the sample. Figure 3(d) shows the Ti 2p spectrum of the substrate. The four peaks of Ti in this spectra are Ti 2p$_{3/2}$ main peak, Ti 2p$_{1/2}$ main peak, Ti 2p$_{3/2}$ satellite peak and Ti 2p$_{1/2}$ satellite peak. The occurance of Ti 2p$_{3/2}$ main peak at 459eV corresponds to Ti$^{+4}$. The +4 state of Ti in STO further confirms the absence of oxygen vacancies in the sample.

Thus, various measurement techniques do not indicate the presence of oxygen vacancies in our structures and therefore the mechanism of oxygen vacancies migration, as put forward for LAO-STO structures, does not appear to be the dominating mechanism for the light enhanced gating effect in LVO-STO heterostructures.

To get further insight of light enhanced gating effect, we performed light intensity dependent sheet resistance measurements for 3 different voltages, \textpm0.6V, \textpm6V and \textpm60V. For these measurements, the samples were kept initially in dark. Then a respective gate voltage was applied along with light illumination at 532nm wavelength for 2 minutes. Figure 4(a) top panel shows the representative examples of resistance evolution in time after turning on the laser light with the intensity of 0.1mW, 10mW and 75mW under negative gate voltage. The percentage resistance change increased from 0.1 to 10mW light intensity, but for 75mW, the percentage change of resistance decreased. This was seen for all negative (-0.6V, -6V and -60V) gate voltages. To elucidate on this, we performed a systemetatic dependence study on laser light intensity ranging from 0.1mW to 75mW. Lower panel on Fig. 4(a) plots the percentage change of resistance on varying the laser light intensity for three back gate voltages. It reveals that the critical laser light intensity (marked with arrow) after which the percentage change of resistance starts decreasing depends on the applied gate voltage; the critical laser light intensity is higher for a higher back gate voltage.

A distinct behavior was observed by application of a positive back gate voltage. Figure 4(b) top panel shows the time evolution of the percentage change of resistance for illuminating the sample with laser light intensity 0.1mW, 10mW and 75mW for positive gate voltages. The summary of systematic measurement for laser light intensity dependence is shown in lower panel of Fig. 4(b) (-$\Delta R$ vs laser power plot). Note, that due to a positive back gate voltage, the resistance of the sample decreases and therefore, the percentage resistance change is negative. This is in contrast to the negative back gate voltage, where resistance increases with increasing light intensity and hence $\Delta R$ is positive. For V$_{g}$=+0.6V, -$\Delta R$ increased continously upto 75mW while for V$_{g}$=+6V, the increase in -$\Delta R$ was not continous, the rate of increase in -$\Delta R$ slowed down after a critical intensity but for V$_{g}$=+60V, -$\Delta R$ saturated as a function of laser intensity after the critical intensity.
  
\section{\label{sec:intro}Discussion}
Our experiment demonstrates a non-trivial interplay between the electrostatic gating effect and simultaneous light illumination in LVO-STO heterostructures.  To approach the understanding of our experimental result we recall that the application of a negative gate voltage depletes electrons at the interface due to the capacitive coupling between the back gate and the LVO-STO interface. Upon shining the laser light, the photo-carriers are excited into the conduction band of the confinement potential of LVO-STO interface. At the applied negative back gate voltage, the carriers should thus be drained out, which would increase the sample resistance. The resistance decrease by manifold is indeed observed in the experiment. However, when the light intensity exceeds a critical value (arrows in Fig. 4(a)), the resistance decreases, which might be thought that the removal of charge carriers by the capacitive coupling becomes less effective and the photo-conductivity effect becomes dominating. By contrast a positive back gate voltage accumulates electrons at the interface due to the same capacitive coupling. The laser light has similar effect; it should increase the number of electrons in the conduction band. Hence, the resistance decreases upon increasing light intensity or gate voltage. However, above some critical value of light intensity, which depends on the gate voltage (arrows in Fig. 4(b)), the resistance saturates. This can be thought that the light cannot promote further electrons with increasing light intensity. These basic phenomena cannot account for all our observations and, in particular, cannot explain the dynamics of photoresponse.  The actual process ought to be by far more complicated. So the light will also be absorbed in LVO layer due to its 1.1eV band gap and this can lead to intertwined processes. The light absorption generates electron-hole pairs, which can be separated by the internal electric field.\cite{Wang2015,Assmann2013} At the same time the light absorption changes the charge balance in LVO layer, which in turn can not only alter the polar catastrophe scenario condition at the interface\cite{Tomar2019} but also possibly the band structure of LVO Mott insulator. Beside this one may expect the modification of conducting channel width, as well as the change in the capacitive coupling between the back-side of the structure and the LVO-STO interface due to the possible light absorption in STO. Despite all those complicated proccesses, that may be happening in LVO-STO stucture under light illumination and application of electric field, our experimental findings our experimental findings provide a possible pathway to tune the electronic state of a conducting oxide interface, which certainly demands a detailed theoretical modelling beyond oxygen vacancy scenario to understand the non-trivial nature of light-gating at LVO-STO interface.

\section{\label{sec:intro}Conclusion}
In conclusion, we have demonstrated the giant conductivity tuning under the illumination of light in presence of electrostatic gating in LVO-STO interface originating even in the absence of detectable oxygen vacancies. This intrinsic phenomenon which might be related to the band filling and can provide an interesting path to tune carrier density in oxide based electronic devices which is otherwise a challenging task. Our observations demand a detailed theoretical modelling to understand the mechanism behind this giant conductivity tuning through "photo-gating" effect to realize oxide based opto-electronic devices.\cite{Assmann2013}

\section{\label{sec:intro}Methods}

\textbf{Substrate preparation.} Thin films of LVO were grown on (001) oriented Ti-terminated STO single crystals. For Ti-termination, method of high temperature annealing followed by DI water etching was employed.\cite{Tomar2017} The STO (001) single crystals were annealed at an optimized temperature of 650$^o$C for two hours in air under ambient conditions. To anneal the substrates, the ramp rate was kept 300$^o$C/hour while heating and it was kept 250 $^o$C/hour while cooling down to room temperature. The annealing accumulated SrO particles on the surface of the substrate which were removed by etching with deionised (DI) water heated at 60$^o$C giving us TiO$_2$ terminated step and terrace like structure. 

\textbf{Thin film growth.} Crystalline and amorphous LVO films were grown on TiO$_2$-terminated STO (001) single crystals using pulsed laser depostion (PLD) system with ceramic LaVO$_4$ as the target. \cite{Hotta2007, Prakash2015} The amorphous LVO films were grown at room temperature and in 1 x 10$^{-6}$ Torr oxygen partial pressure and crystalline LVO films were grown at substrate temperature of 600$^o$C and oxygen partial pressure of 1 x 10$^{-6}$ Torr. The film thickness was monitored using the reflection high-energy electron diffraction (RHEED) technique.

\textbf{Transport and photoresponse measurements.} The electrical transport as well as the photoresponse measurements of the grown heterostructures were done using physical property measurement system (PPMS) by Quantum Design. For these measurements, contacts were made by ultrasonically wire-bonding the interface. For electrostatic gating, Kiethley source meters were used. Also, Diode-pumped solid state (DPSS) lasers having wavelength 405 nm and 532 nm were used to check the effect of photo-excitation of the carriers in our oxide samples. 
The XRD study of the sample under various stimuli was done using Bruker thin film X-ray diffractometer Discover-8 and Raman spectroscopic studies were done using WITec Confocal Raman Spectrometer. Hard X-ray (HX) photoemission measurements were carried out on the samples at BL-47XU of SPring-8.

$\textbf{Acknowledgements}$: 
SC acknowledges financial support from Department of Science and Technology (DST) India, Nano Mission project number (SR/NM/NS-1007/2015). The experiments at SPring-8 were performed with the approval of the Japan Synchrotron Radiation Research Institute (JASRI) (Proposal No.2016A1210).

 $\textbf{Data Availability}$: The data that support the findings of this study are available from the corresponding author upon reasonable request.



\end{document}